\documentclass[epj]{svjour}   

\usepackage{times}
\usepackage{amsmath}

\input epsf

\numberwithin{equation}{section} 
\begin{document}

\title{Universal crossing probabilities and incipient 
spanning clusters in directed percolation}

\author{L. Turban\thanks{\email{turban@lpm.u-nancy.fr}}}                    

\institute{Laboratoire de Physique des Mat\'eriaux\thanks{UMR CNRS 7556}, 
Universit\'e  Henri Poincar\'e (Nancy I), BP 239,\\ 
54506 Vand\oe uvre l\`es Nancy Cedex, France}                                

\date{Received 10 February 2003} 

\abstract{
Shape-dependent universal crossing probabilities are studied, {\it via} 
Monte Carlo simulations, for bond and site directed percolation 
on the square lattice in the diagonal direction, at the percolation 
threshold. In a dynamical interpretation, the crossing probability
is the probability that, on a system with size $L$, an epidemic spreading without 
immunization remains active at time $t$. Since the system is strongly anisotropic, 
the shape dependence in space-time enters through the effective aspect ratio 
$r_{\rm eff}=ct/L^z$, where $c$ is a non-universal constant 
and $z$ the anisotropy expo\-nent. A particular attention is paid to the influence of 
the initial state on the universal behaviour of the crossing probability. Using 
anisotropic finite-size scaling and generalizing a simple argument given by Aizenman 
for isotropic percolation, we also obtain the behaviour of 
the probability to find $n$ incipient spanning clusters on a finite system 
at time $t$. The numerical results are in good agreement with the conjecture.
\PACS{{64.60.Ak}{Renormalization-group, fractal, and 
percolation studies of phase transitions}
\and{05.50.+q}{Lattice theory and statistics (Ising, Potts, etc.)}  
\and{02.50.-r}{Probability theory, stochastic processes, and statistics}} 
} 

\maketitle

\section{Introduction}
\label{sec:1}
 
Most of the properties of the critical state at a second order phase transition
are related to the existence of self-similar correlated clusters. In the 
percolation problem, the critical point is characterized by the spontaneous 
occurence of infinite clusters of connected sites, crossing the system in the 
thermodynamic limit.

The study of crossing probabilities for standard percolation on finite systems at 
the percolation threshold has been an active field of research during the last 
decade~[1--44] (for a recent review see~\cite{cardy01}). 

In two dimensions, the crossing probability $\pi$ can be defined as the 
probability to find at least one cluster joining two opposite edges of 
a rectangular-shaped system with length~$L_\parallel$ and width~$L_\perp$. At the 
percolation threshold, in the finite-size scaling (FSS) limit, {\it i.e.}, when 
$L_\parallel\to\infty$, $L_\perp\to\infty$, while the aspect ratio 
$r=L_\parallel/L_\perp$ remains fixed, the crossing probability is a {\em 
scale-invariant universal function} $\pi(r)$ of the aspect 
ratio~\cite{langlands92,cardy92}.   

Following the extensive numerical study of Langlands {\it et 
al.}~\cite{langlands92}, an analytical expression for $\pi(r)$ at criticality was 
derived by Cardy~\cite{cardy92}. 
Using the relation between percolation and the cluster formulation of the $q$-state 
Potts model in the limit $q\to1 $~\cite{kasteleyn69,kasteleyn72} as well as the 
techniques of boundary conformal field theory, he was able to calculate the 
probability for two non-overlapping segments on the edge of the half-plane to be 
connected. The crossing probability in the rectangular geometry could then be 
deduced from the appropriate conformal mapping. One may notice that the {\em scale 
invariance} of $\pi(r)$ is a non-trivial property, resulting from the occurence of 
a vanishing scaling dimension $x(q)$ associated with a boundary condition changing 
operator of the Potts model in the limit $q\to 1$ for percolation. 

Other exact crossing formulas were later obtained for different 
geometries~\cite{pinson94,watts96}. Since then, mathematically rigorous 
proofs have been provided for some of these results~\cite{smirnov01,schramm01}. 

Crossing probabilities on same-spin Ising 
clusters~\cite{langlands00,lapalme01,arguin02a}
and Potts clusters~\cite{arguin02b} in two dimensions have been also considered. 

Recently, some attention has been paid to the properties of {\em incipient 
spanning clusters}~[31--44]. The probability to find $n$ incipient spanning 
clusters, connecting two disjoint segments at the boundary of a finite 
critical system, has been studied through Monte Carlo 
simulations~[31--39]. Rigourous bounds on the spanning 
probability~\cite{aizenman97,aizenman98} have been obtained as 
well as analytical expressions through conformal and Coulomb-gas 
methods~\cite{cardy98,aizenman99,cardy02}. 

In two recent works~\cite{turban02,turban03}, we started a study of the critical 
crossing probability (in space-time) for off-equilibrium systems in 1+1 dimensions.
In these strongly anisotropic systems, when the temperature-like scaling field 
$\Delta$ is non-vanishing, the correlation functions generally display an 
exponential decay, in the space and time directions, characterized by a correlation 
length $\xi$ and a relaxation time $\tau$ diverging respectively as~\cite{binder89}
\begin{equation}
\xi=\hat{\xi}\Delta^{-\nu}\,,\qquad \tau=\hat{\tau}\Delta^{-z\nu}\,,
\label{e1.1:xitau}
\end{equation}
when $\Delta\to0$. The prefactors $\hat{\xi}$ and $\hat{\tau}$ are non-universal 
amplitudes and $z$ is the dynamical (anisotropy) exponent. 

The directed percolation problem~\cite{broadbent57,kinzel83,hinrichsen00} was 
studied numerically through Monte Carlo simulations in~\cite{turban02} and 
analytical results were obtained for an exactly solvable diffusion-limited 
coagulation process in~\cite{turban03}. 
In these sytems, the crossing probability in the time direction $\pi_t$ is defined 
as the probability that the system of size~$L$ remains active at time~$t$. 
Ani\-so\-tro\-pic scaling~\cite{binder89,hucht02} then implies that the appropriate 
aspect ratio is $r=t/L^z$, {\it i.e.}, the {\em rescaled time}. We found that in the 
FSS limit, as for isotropic critical systems, the critical crossing 
probability is a scale-invariant universal function of an {\em effective 
aspect ratio} which is the product of $r$ by a non-universal constant $c$~\cite{r}.

In the present work, we continue the examination of the universal properties of the 
crossing probability $\pi_t$ in directed percolation which, in a dynamical 
interpretation, essentially corresponds to the {\em survival probability of an 
epidemic spreading without immunization}~\cite{harris74}. Section~\ref{sec:2} is 
devoted to a detailed study of the influence of the initial state on the universal 
behaviour of $\pi_t$. In Section~\ref{sec:3} we present some results concerning the 
scaling behaviour of the probability to find $n$ incipient spanning clusters, 
{\it i.e.}, {\em $n$ disjoint centres of infection} surviving in the FSS limit.

\section{Critical crossing probability}
\label{sec:2}

\begin{figure}[t]
\epsfxsize=7cm
\begin{center}
\vglue0.mm
\hspace*{-2.mm}\mbox{\epsfbox{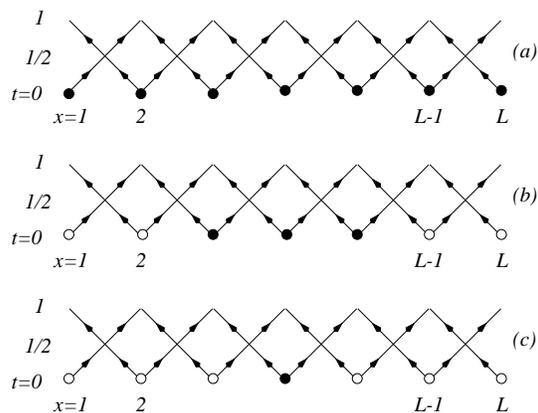}}
\end{center}
\caption{Directed percolation on the square lattice in the diagonal direction. 
There are $L$ sites in the spatial direction which are either occupied $(\bullet)$ 
or empty $(\circ)$. Three types of initial states are considered with a) $L$ 
occupied sites, b) $l<L$ nearest-neighbour occupied sites and c) a single occupied 
site. In the last two cases, for FBC, the occupied sites are located in the middle 
of the system.}  
\label{fig1_dp}  
\end{figure}

We study the critical crossing probability $P_t$, in the time direction, 
for bond and site directed percolation
in $1+1$ dimensions through Monte Carlo 
simulations on the square lattice. The time axis is oriented in the diagonal 
direction. The space and time coordinates take alternatively integer and 
half-integer values on successive spatial rows. On a site located at $(x,t)$, 
with $1\leq x\leq L$, two directed bonds are leaving, which terminate on the 
nearest-neighbour sites at $(x\pm1/2,t+1/2)$ as shown in Figure~\ref{fig1_dp}. 
We use either free boundary conditions (FBC) or periodic boundary conditions 
(PBC) in the spatial direction. In the later case, $x$ is defined modulo $L$.

In the epidemic language, occupied (empty) sites correspond to infected (healthy) 
individuals. In the bond problem, the bonds are independently open with probability 
$p$. A site is occupied at $t+1/2$ when it is connected {\it via} an open bond to a site 
which was occupied at time $t$. In the site problem all the bonds are open and a 
site is occupied with probability $p$ at time $t+1/2$ when at least one of its 
first neighbours at time $t$ is occupied. 

A directed percolation cluster is a collection of connected occupied sites starting 
from some source at $t=0$. We consider successively the three types of initial 
states shown in Figure~\ref{fig1_dp}. A sample contributes to the crossing 
probability when at least one cluster survives at time $t$. The simulations are 
performed at the percolation threshold, $p_{\rm c}^{\rm bond}=0.644700185(5)$ 
for the bond problem and $p_{\rm c}^{\rm site}=0.70548522(4)$ for the site 
problem. These values, as well as the dynamical exponent $z=1.580745(10)$, are 
taken from~\cite{jensen99}. The crossing probability $P_t(L,t,l)$ is 
determined as a function of the aspect ratio, $r=t/L^z$, and the fraction 
$f=l/L$ of occupied sites in the initial state, using $10^6$ 
samples with sizes up to $L=2^{10}$. When $f$ is non-vanishing, $P_t$ converges 
to a scale-invariant function, $\pi_t(r,f)$, in the FSS 
limit where $L$, $t$ and $l\to\infty$, while keeping $r$ and $f$ fixed.

\subsection{Initial state with a fully occupied lattice}
\label{sec:2.1}

\begin{figure}[t]
\epsfxsize=8cm
\begin{center}
\vglue0.mm
\hspace*{-2.mm}\mbox{\epsfbox{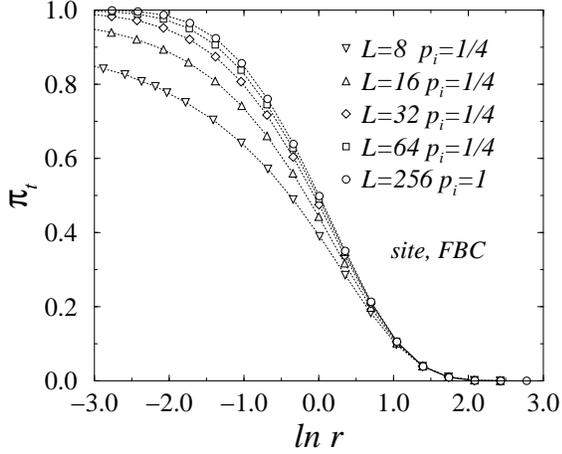}}
\end{center}
\vglue-3mm
\caption{Crossing probability as a function of $\ln r$ for site percolation 
with FBC. With increasing size, the values of $\pi_t$ obtained when the $L$ 
sites are randomly occupied in the initial state with probability $p_i=1/4$, 
converge to the values obtained for $p_i=1$ on a lattice with size $L=256$. The 
statistical errors are smaller than the symbols.}  
\label{fig2_dp}  
\end{figure}

When the $L$ sites are randomly occupied with probability $p_i>0$, the crossing 
probability converges to a scale-invariant function in the FSS limit.
This is illustrated in Figure~\ref{fig2_dp} in the case of site percolation with 
FBC. When $p_i=1$ the convergence is from above and the results obtained with 
$L=8$ are already quite close to the FSS limit. Thus $p_i$ is an irrelevant 
variable when $f>0$. In the following we always take $p_i=1$ in order to obtain 
a better convergence towards the scale-invariant behaviour. 
   
\begin{figure}[t]
\epsfxsize=8cm
\begin{center}
\vglue0.mm
\hspace*{-2.mm}\mbox{\epsfbox{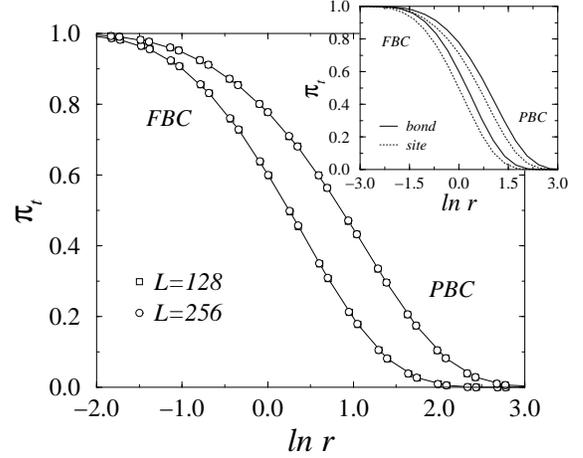}}
\end{center}
\vglue-3mm
\caption{Crossing probability as a function of $\ln r$ with a fully 
occupied lattice in the initial state. The Monte Carlo results for the largest 
size ($L=256$) are shown in the inset for the bond and site problems with FBC 
and PBC. The main figure shows the data collapse on two different universal 
curves for FBC and PBC in the scaling limit, after shifting the data for site 
percolation, as explained in the text. The statistical errors are smaller than 
the symbols.}  
\label{fig3_dp}  
\end{figure}

We now consider site and bond directed percolation with either FBC or PBC in 
order to study the universality of the crossing probability. The lattice is  
fully occupied in the initial state. The behaviour of the crossing probability as a 
fonction of the aspect ratio is shown in Figure~\ref{fig3_dp}. The raw data 
converge quickly to scale-invariant functions as the system size is increased. 
They are shown in the inset for the largest size, $L=256$. Different functions 
are obtained for the two types of boundary conditions and the curves for the 
bond and site problems are shifted with respect to one another, along the 
horizontal axis, by an amount which is the same for FBC and PBC. The main figure 
shows the data collapse on universal curves, one for each type of boundary 
condition, after the results for site percolation have been shifted.

In order to explain this behaviour let us consider 
the scaling behaviour of the crossing probability for an off-critical system 
with a deviation $\Delta=|p-p_{\rm c}|$ from the percolation threshold. Under an 
anisotropic change of the length scale by a factor~$b$ the length transforms as 
$L'=L/b$, the time as $t'=t/b^z$ and $\Delta$ as $\Delta'=b^{1/\nu}\Delta$, 
where $\nu$ is the correlation length exponent defined in~(\ref{e1.1:xitau}). 
Assuming that, as indicated by the Monte Carlo data, the crossing probability is 
scale invariant, then it satisfies the scaling relation
\begin{equation}
P_t(L,t,\Delta)=P_t\left(\frac{L}{b},\frac{t}{b^z},b^{1/\nu}\Delta\right)\,.
\label{e2.1:scal1}
\end{equation}
With $b=\Delta^{-\nu}$ one obtains
\begin{equation}
P_t(L,t,\Delta)=
P_t\left(\frac{L}{\Delta^{-\nu} },\frac{t}{\Delta^{-z\nu}},1\right)
=g\left(\frac{L}{\xi},\frac{t}{\tau}\right)\,,
\label{e2.1:scal2}
\end{equation}
where $g(u,v)$ is a universal fonction of its dimensionless arguments.   
The FSS limit corresponds to $\Delta=0$ and $b=L$ 
in~(\ref{e2.1:scal1}), which gives
\begin{equation}
P_t(L,t,0)=P_t\left(1,\frac{t}{L^z},0\right)=\pi_t(cr)\,.
\label{e2.1:scal3}
\end{equation}
For each type of boundary conditions, the crossing probability is a scale-invariant 
uni\-ver\-sal func\-tion of the effective aspect ratio $r_{\rm eff}=cr$. The 
non-universal amplitude $c$ depends on the choice of the length 
and time units. It can be expressed as a function of $\hat{\xi}$ and 
$\hat{\tau}$~\cite{hucht02}. Comparing~(\ref{e2.1:scal2}) 
and~(\ref{e2.1:scal3}), $r$ appears through the dimensionless ratio $v/u^z$ so 
that
\begin{equation}
r_{\rm eff}=\frac{t/\tau}{(L/\xi)^z}=\frac{\hat{\xi}^z}{\hat{\tau}}r\,,\qquad
c=\frac{\hat{\xi}^z}{\hat{\tau}}\,.
\label{e2.1:c}
\end{equation}
The non-universal constant $c$ is different for the two percolation problems. 
According to~(\ref{e2.1:scal3}), for a given type of boundary condition, the 
crossing probabilities $\pi_t$ are identical for the site and bond problems when 
the values of the aspect ratio, $r_{\rm s}$ and~$r_{\rm b}$, satisfy the 
relation $c_{\rm s}r_{\rm s}=c_{\rm b}r_{\rm b}$. Thus the data collapse in 
Figure~\ref{fig3_dp} is obtained through a shift of the site percolation data by
\begin{equation}
\delta^r_{\rm sb}=\ln r_{\rm b}-\ln r_{\rm s}=
\ln\left(\frac{c_{\rm s}}{c_{\rm b}}\right)=.2476(8)\,.
\label{e2.1:shift}
\end{equation}
The numerical value of $\delta^r_{\rm sb}$ was estimated in~\cite{turban02} 
through a least-square fit. 

The crossing probability can be calculated using a transfer matrix in the time 
direction \tens{T}, working with a restricted basis of states corresponding to 
configurations with at least one occupied site~\cite{derrida80}. The matrix element 
$T_{\alpha\beta}$ gives the probability that the configuration in state 
$|\alpha\rangle$ at time $t$ leads to the configuration in state  
$|\beta\rangle$ at time $t+1$. Given an initial state $|i\rangle$, the 
crossing probability then reads
\begin{equation}
\pi_t(cr)=\sum_\alpha\langle i|\tens{T}^t|\alpha\rangle\,.
\label{e2.1:tm1}
\end{equation}
When $r=t/L^z\gg1$, the leading contribution comes from the largest 
eigenvalue $\Lambda_{\rm max}<1$ of $\tens{T}$ and
\begin{equation}
\pi_t(cr)\sim\Lambda_{\rm max}^t=\exp\left(-a\frac{ct}{L^z}\right)
=\exp(-acr) 
\label{e2.1:tm2}
\end{equation}
\begin{figure}[t]
\epsfxsize=8cm
\begin{center}
\vglue0.mm
\hspace*{-2.mm}\mbox{\epsfbox{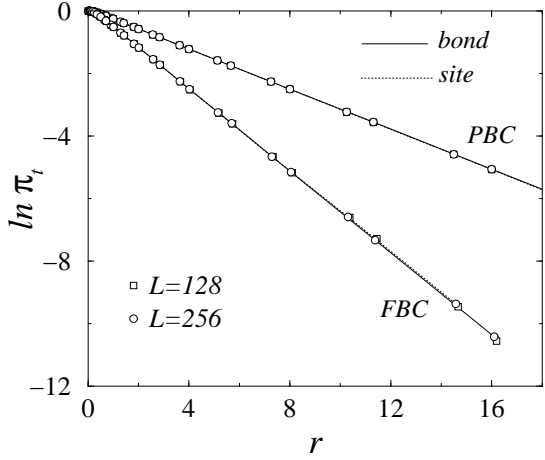}}
\end{center}
\vglue-3mm
\caption{Logarithm of the crossing probability as a function of $r$ with a 
fully occupied lattice in the initial state. For the site problem, $r$ has 
been multiplied by $c_{\rm s}/c_{\rm b}$. In the scaling limit, a linear 
universal behaviour is obtained, with different slopes for FBC and PBC.}  
\label{fig4_dp}  
\end{figure}
since the relaxation time $|\ln(\Lambda_{\rm max})|^{-1}$ scales as $L^z$. In 
the argument of the exponential, $a$ is some universal constant. Thus $\ln\pi_t$ 
is a universal linear function of $cr$ when $r\gg1$. This universal 
linear variation is shown in Figure~\ref{fig4_dp} where $r$ for the site 
problem has been multiplied by $c_{\rm s}/c_{\rm b}=1.281(1)$ in order to 
eliminate the influence of the non-universal constant $c$ on the data collapse.
Since the argument leading to~(\ref{e2.1:tm2}) is quite general, the same 
behaviour is expected for the other types of initial states considered below.

\subsection{Initial state with a sequence of occupied sites}
\label{sec:2.2}

\begin{figure}[t]
\epsfxsize=8cm
\begin{center}
\vglue0.mm
\hspace*{-2.mm}\mbox{\epsfbox{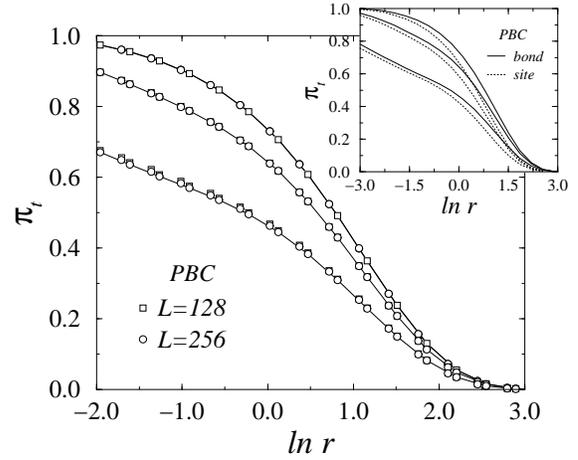}}
\end{center}
\vglue-3mm
\caption{Crossing probability as a function of $\ln r$ with a fraction 
$f=l/L$ of occupied lattice in the initial state and PBC. The Monte Carlo 
results for the largest size ($L=256$) are shown in the inset for the bond and 
site problems with $f=1/2$, $1/4$ and $1/16$ from top to bottom. The main figure 
shows the data collapse in the scaling limit on three different universal 
curves, one for each value of $f$, when the data for site percolation are 
appropriately shifted. The statistical errors are smaller than the symbols.}  
\label{fig5_dp}  
\end{figure}

\begin{figure}[htb]
\epsfxsize=8cm
\begin{center}
\vglue0.mm
\hspace*{-2.mm}\mbox{\epsfbox{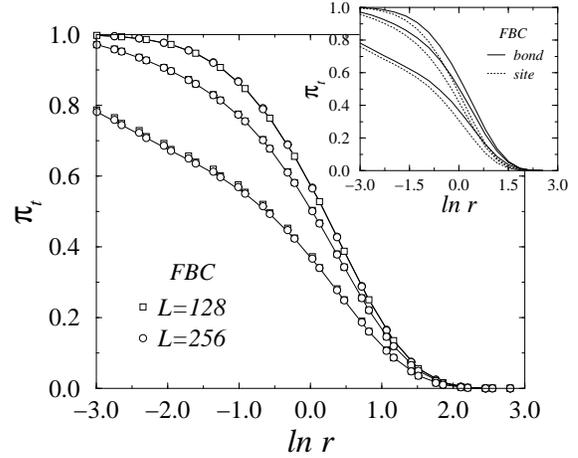}}
\end{center}
\vglue-3mm
\caption{As in Figure~\protect\ref{fig5_dp} for FBC.}  
\label{fig6_dp}  
\end{figure}

Let us consider now the behaviour of the crossing probability when $l$ 
consecutive sites are occupied in the initial state, as shown in 
Figure~\ref{fig1_dp}b. The results for PBC and FBC are presented in 
Figures~\ref{fig5_dp} and \ref{fig6_dp}, respectively, for three values of the 
occupation ratio $f=l/L$, from top to bottom $f=1/2$, $1/4$, $1/16$. The raw 
data for the largest size, $L=256$, are shown in the insets. As above, different 
curves are obtained for the site and bond problems. Shifting the site results 
by~$\delta^r_{\rm sb}$, the data collapse on three universal curves, one for 
each value of $f$, as shown on the main figures. 

The problem involves the new length scale, $l$, and the crossing probability 
scales as
\begin{equation}
P_t(L,t,l,\Delta)=P_t\left(\frac{L}{b},\frac{t}{b^z},\frac{l}{b},b^{1/\nu}\Delta
\right)\,.
\label{e2.2:scal1}
\end{equation}
With $b=\Delta^{-\nu}$ one obtains
\begin{equation}
P_t(L,t,l,\Delta)=
P_t\!\!\left(\frac{L}{\Delta^{-\nu}},\frac{t}{\Delta^{-z\nu}},
\frac{l}{\Delta^{-\nu}},1\right)\!\!
=\!h\!\left(\frac{L}{\xi},\frac{t}{\tau},\frac{l}{\xi}\right),
\label{e2.2:scal2}
\end{equation}
where $h(u,v,w)$ is a universal function of its arguments. In the 
FSS limit, at the critical point, equation~(\ref{e2.2:scal1}) 
gives
\begin{equation}
P_t(L,t,l,0)=P_t\left(1,\frac{t}{L^z},\frac{l}{L},0\right)
=\pi_t(cr,f)\,.
\label{e2.2:scal3}
\end{equation}
The dimensionless occupation ratio $f$ does not introduce any new 
non-universal constant in the scaling limit where $L$ and $l\to\infty$.
Thus, as shown in Figures~\ref{fig5_dp} and~\ref{fig6_dp}, the crossing 
probability is a scale-invariant universal function of the effective aspect 
ratio $r_{\rm eff}$, which depends on the value of $f$ as well as on the boundary 
conditions. One may notice a slightly slower convergence to the limiting 
behaviour when $f$ decreases.

\subsection{Initial state with a single occupied site}
\label{sec:2.3}

\begin{figure}[tbh]
\epsfxsize=8cm
\begin{center}
\vglue0.mm
\hspace*{-2.mm}\mbox{\epsfbox{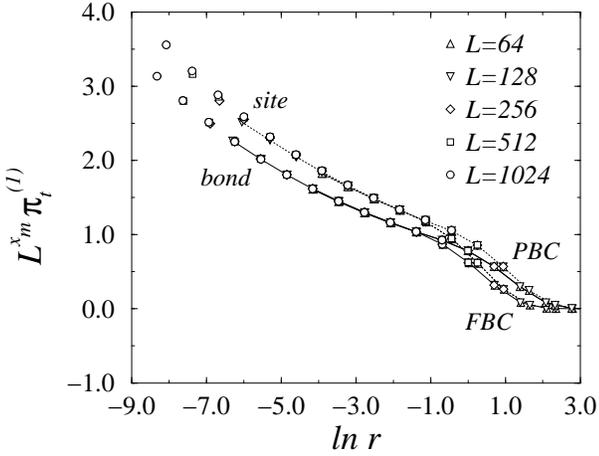}}
\end{center}
\vglue-3mm
\caption{Crossing probability for site (dotted line) and bond (full line) directed 
percolation with a single occupied site in the initial state as a function of $\ln 
r$. After rescaling $\pi_t^{(1)}$ by a factor $L^{x_m}$, the size effects are 
eliminated. Although the data for the site problem have been shifted by 
$\delta^r_{\rm sb}$, a different behaviour is still obtained for site and bond 
percolation.}  
\label{fig7_dp}  
\end{figure}

When a single site (located in the middle of the system for FBC) is occupied in the 
initial state, the crossing probability $\pi_t^{(1)}$ is no longer scale 
invariant. It acquires a dimension which is the scaling 
dimension of the order parameter, 
\begin{equation}
x_m=\frac{\beta}{\nu}=.252072(8)\,, 
\label{e2.3:xm}
\end{equation}
corresponding to the values 
\begin{equation}
\beta=.276486(8)\,,\qquad\nu=1.096854(4)\,, 
\label{e2.3:betanu}
\end{equation}
taken from~\cite{jensen99}. This behaviour follows from the scaling of $P_t^{(1)}$, 
the probability that a cluster grown from a single central occupied site survives 
after $t$ time steps,
\begin{equation}
P_t^{(1)}(L,t,\Delta)
=b^{-x_m}P_t^{(1)}\left(\frac{L}{b},\frac{t}{b^z},b^{1/\nu}\Delta\right)\,.
\label{e2.3:scal1}
\end{equation}
On the infinite system with $b=t^{1/z}$, one obtains  
\begin{equation}
P_t^{(1)}(\infty,t,\Delta)
=t^{-x_m/z}P_t^{(1)}\left(\infty,1,t^{1/z\nu}\Delta\right)\,,
\label{e2.3:scal2}
\end{equation}
in agreement with equation~(94) in reference~\cite{hinrichsen00}, where the 
exponent $\beta'/\nu_\parallel=\beta/\nu_\parallel=x_m/z$ for directed percolation 
with our notations. Taking $b=L$ at criticality ($\Delta=0$) leads in the FSS limit 
to 
\begin{equation}
P_t^{(1)}(L,t,0)
=L^{-x_m}P_t^{(1)}\left(1,\frac{t}{L^z},0\right)=\pi_t^{(1)}(cr,L)\,,
\label{e2.3:scal3}
\end{equation}
As shown in Figure~\ref{fig7_dp} a good data collapse is 
obtained for $L^{x_m}\pi_t^{(1)}$ with the different system sizes, $L=2^6$ to 
$2^{10}$. In this figure, the data for the site percolation problem have been 
shifted by $\delta_{\rm sb}$ given in~(\ref{e2.1:shift}) in order to take into 
account the non-universal constants $c_{\rm s}$ and $c_{\rm b}$. 

\begin{figure}[t]
\epsfxsize=8cm
\begin{center}
\vglue0.mm
\hspace*{-2.mm}\mbox{\epsfbox{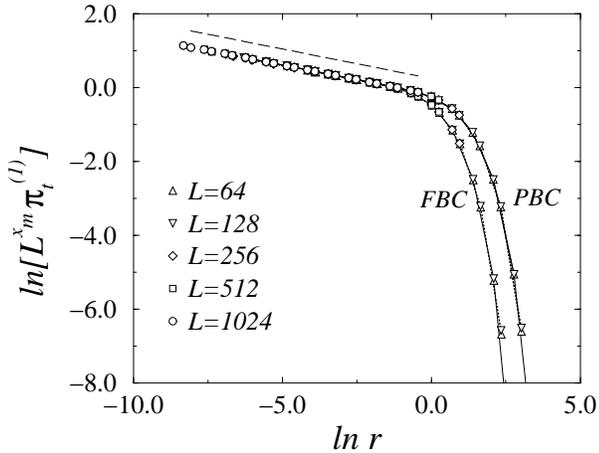}}
\end{center}
\vglue-3mm
\caption{Once $\ln(L^{x_m}\pi_t^{(1)})$ for the site problem has been also shifted 
by $\delta^\pi_{\rm sb}$ in order to eliminate the influence of the non-universal 
amplitude in~(\protect\ref{e2.3:scal5}), the data collapse on two universal curves  
for FBC and PBC, respectively. For small values of $\ln r$ a linear variation is 
obtained with a slope $-x_m/z$, as indicated by the dashed line.}  
\label{fig8_dp}  
\end{figure}

For small values of $r$, $L^z\gg t$, the boundary conditions are irrelevant 
since, with a single occupied site in the initial state, the crossing cluster 
issued from the middle of the sample does not explore the boundary regions. For the 
same reason $\pi_t^{(1)}$ is independent of the size $L$ in this regime. 
Different curves are obtained for the site and bond problems for a given type of 
boundary conditions, meaning that a new non-universal constant is involved. In a 
log-log plot, it is clear that a single curve can be obtained through a vertical 
shift $\delta^\pi_{\rm sb}$ of the data for the site problem. The resulting data 
collapse is shown in Figure~\ref{fig8_dp}. Thus the new non-universal constant 
$A$ is a multiplicative one, either $A_{\rm s}$ or $A_{\rm b}$, and the shift is 
given by: 
\begin{equation}
\delta^\pi_{\rm sb}=\ln\left(\frac{A_{\rm b}}{A_{\rm s}}\right)=-.178(4)\,.
\label{e2.3:shift}
\end{equation}
It was calculated by first shifting the data of the site problem by 
$\delta^r_{\rm sb}$. Due to this horizontal shift of the site data, a direct 
calculation of the vertical shift is not possible. Thus, considering 
$y=\ln(L^{x_m}\pi_t^{(1)})$ as a function $\ln r$, the data of the bond problem 
were fitted to a quadratic polynomial in the linear region, for small values of 
$\ln r$. Then $\delta^\pi_{\rm sb}$ was obtained as the average of $y_{\rm 
b}-y_{\rm s}$, evaluated at small enough values of $\ln r$ for the 
site problem. 

Alternatively, the scaling behaviour of $\pi_t^{(1)}$ in~(\ref{e2.3:scal3}) 
can be recovered by assuming that, asymptotically when $f\ll1$, the scale-invariant 
crossing probability in~(\ref{e2.2:scal3}) behaves as
\begin{equation}
\lim_{f\ll1}\pi_t(cr,f)=f^{x_m}\varphi(cr)
=\left(\frac{l}{L}\right)^{x_m}\varphi(cr)\,,
\label{e2.3:scal4}
\end{equation}
so that $L^{x_m}\pi_t^{(1)}=l^{x_m}\varphi(cr)$.

Furthemore the non-universal amplitude, $A_{\rm s}$ or $A_{\rm b}$, which 
appears only in $\pi_t^{(1)}$, {\it i.e.}, when the number $l$ of occupied sites in the 
initial state is O(1), can be explained if one associates to these $l$ sites an 
effective length $l_{\rm eff}$ given by
\begin{equation}
l_{\rm eff}=l+\delta l\,,\qquad\delta l=\delta l_{\rm s},\delta l_{\rm b}\,,
\label{e2.3:leff}
\end{equation}
where the increment $\delta l$ is O(1) and non-universal. Thus, replacing $f$ by 
$f_{\rm eff}=(l+\delta l)/L$, there is no change in the FSS limit when $f>0$, 
but when $l=1$ we obtain 
\begin{equation}
L^{x_m}\pi_t^{(1)}=(1+\delta l)^{x_m}\varphi(cr) 
\label{e2.3:scal5}
\end{equation}
with a non-universal amplitude given by 
\begin{equation}
A=(1+\delta l)^{x_m}\,.
\label{e2.3:ampl}
\end{equation}
According to~(\ref{e2.3:shift}) we have $\delta l_{\rm s}>\delta l_{\rm b}$.

Since, as noticed above, $\pi_t^{(1)}$ becomes independent of $L$ for small 
values of $r$, the scaling function $\varphi(w)$ with $w=cr$ has to behave 
as $w^{-x_m/z}$ when $w\ll1$. This leads to a slope 
$-x_m/z=-\beta/z\nu=-.159464(5)$ for the linear part of 
$\ln(L^{x_m}\pi_t^{(1)})$ as a function of $\ln r$, in good agreement with 
the numerical results in~Figure~\ref{fig8_dp}.

\section{Critical spanning clusters}
\label{sec:3}

\begin{figure}[t]
\epsfxsize=8cm
\begin{center}
\vglue0.mm
\hspace*{-2.mm}\mbox{\epsfbox{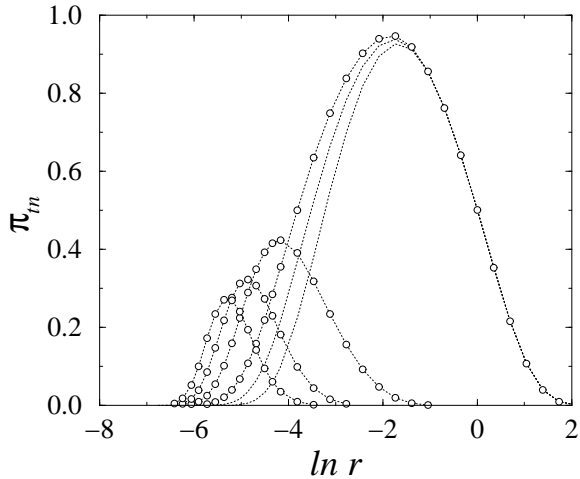}}
\end{center}
\vglue-3mm
\caption{Probability $\pi_{tn}$ to find $n$ critical spanning clusters in the time 
direction for the site problem with FBC when $L=256$. The number of clusters is 
$n=1$ to~$4$ from right to left. For $n=1$, the spanning probabilities obtained 
when $L=64$ and $128$ are also shown.}  
\label{fig9_dp}  
\end{figure}

In this Section we present a preliminary study of the scaling behaviour of the 
probability $\pi_{tn}(r)$ to find exactly $n$ independent clusters surviving after 
$t=rL^z$ time steps on a system with size $L$ in the FSS limit. In the epidemic 
interpretation $\pi_{tn}$ is the probability to find $n$ disjoint centres of 
infection surviving in the FSS limit. This problem presents some similarity with the 
statistics of family names first studied by Galton and Watson~\cite{watson74} (see 
also~\cite{broeker99,janssen01}).

The spanning probability was estimated by generating $10^6$ samples with 
$L=64,128$ and $256$ for the site problem with FBC or PBC and $L$ occupied sites in 
the initial state. As the size increases, it converges to a scale-invariant 
function $\pi_{tn}$ of the effective aspect ratio $r_{\rm eff}$. A better 
convergence is obtained with FBC, thus we shall only analyse the results obtained 
with this type of boundary conditions. An example of the raw data for $\pi_{tn}$ at 
the largest sizes studied is shown in Figure~\ref{fig9_dp}. 

The scaling behaviour can be obtained by extending to the case of a strongly 
anisotropic system a simple argument proposed earlier for isotropic 
percolation~\cite{aizenman97,cardy98}. 

\begin{figure}[t]
\epsfxsize=7cm
\begin{center}
\vglue0.mm
\hspace*{-2.mm}\mbox{\epsfbox{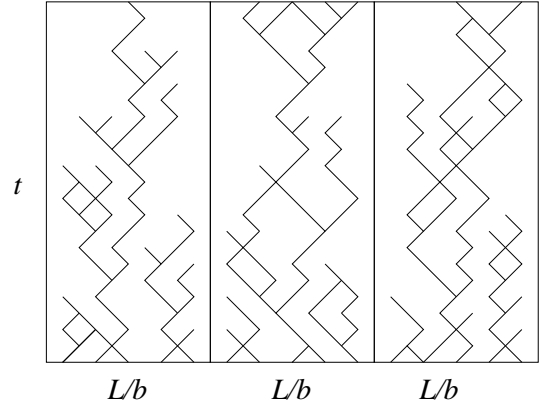}}
\end{center}
\caption{$n$ spanning clusters on a system with size $L$ at time $t$ in $1+1$ 
dimensions are considered to result typically from the juxtaposition of $n/b$ 
spanning clusters on $b$ subsystems with size $L/b$.}  
\label{fig10_dp}  
\end{figure}

Let $P_t(n,r)$ denote the probability to have exactly $n$ spanning clusters at time 
$t$, on a system in $d+1$ dimensions, with volume $L^d$. The scale invariance of 
$P_t$ implies that the size dependence is wholly contained in the aspect ratio $r$. 
Dividing the system in $b^d$ subsystems with the same volume $(L/b)^d$, as shown 
schematically in Figure~\ref{fig10_dp} for $d=1$ and $b=n$, the dominant event will 
correspond to $n'=n/b^d$ spanning clusters in space-time for each subsytem. Since 
the aspect ratio of the large system is $r=t/L^z$, it becomes $r'=t/(L/b)^z=rb^z$ 
in the smaller subsystems. Thus, up to prefactors, one expects the following 
behaviour:
\begin{equation}
P_t(n,r)\sim
\left[P_t(b^{-d}n,b^zr)\right]^{b^d}\,.
\label{e3:scal1}
\end{equation}
According to the transfer-matrix argument of Section~\ref{sec:2.1}, a leading 
exponential dependence on $t$ (through $r$) is expected. Thus, with $b=n^{1/d}$, we 
obtain
\begin{equation}
P_t(n,r)\sim\left[P_t(1,n^{z/d}r)\right]^n
\sim\exp(-\alpha rn^\omega)\,.
\label{e3:scal2}
\end{equation}
Applying the first relation to the exponential form allows us to identify the 
exponent $\omega=1+z/d$ so that the scale-invariant crossing probability behaves as
\begin{equation} 
\pi_{tn}(r)\sim\exp(-\alpha rn^{1+z/d})\,,
\label{e3:scal3}
\end{equation}
where $\alpha$ is a non-universal prefactor. Actually, the cros\-sing 
pro\-ba\-bi\-li\-ty $\pi_{tn}$, like $\pi_t$, is a universal function 
of the effective aspect ratio $cr$. This scaling behaviour
generalizes the form conjectured by Aizenman for isotropic percolation in $D=d+1$ 
dimensions~\cite{aizenman97}, which was recently tested numerically with 
success~\cite{shchur02}. The isotropic limit corresponds to $z=1$ 
and $d=D-1$ in~(\ref{e3:scal3}). The anisotropic result should remain valid 
below the upper critical dimension $d_{\rm c}=4$ for directed 
percolation~\cite{kinzel83}. At and above $d_{\rm c}$, as in the isotropic 
case, one expects a proliferation of spanning clusters, linked with 
the breakdown of hyperscaling~\cite{aizenman97,aharony84}.

\begin{figure}[t]
\epsfxsize=8cm
\begin{center}
\vglue0.mm
\hspace*{-2.mm}\mbox{\epsfbox{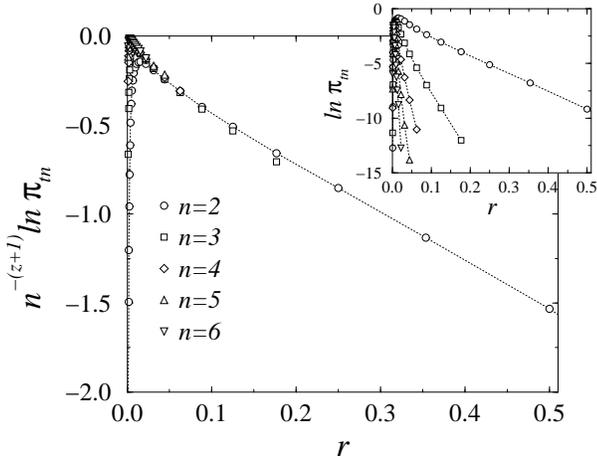}}
\end{center}
\vglue-3mm
\caption{Scaling behaviour of the crossing probability $\pi_{tn}(r)$ for the site 
percolation problem with FBC when $L=256$. The inset shows the linear variation of 
$\ln\pi_{tn}$ versus $r$ with a slope depending on $n=2$ to~$6$ from top to bottom. 
A good data collapse is obtained in the main figure, in agreement 
with~(\protect\ref{e3:scal3}) for $d=1$, when $\ln\pi_{tn}$ is rescaled by 
$n^{-(z+1)}$.}  
\label{fig11_dp}  
\end{figure}

\begin{figure}[tbh]
\epsfxsize=8cm
\begin{center}
\vglue0.mm
\hspace*{-2.mm}\mbox{\epsfbox{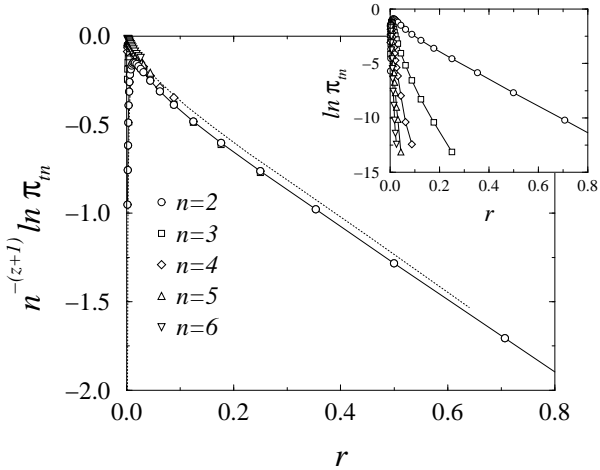}}
\end{center}
\vglue-3mm
\caption{As in Figure~\protect\ref{fig11_dp} for the bond percolation problem.
The dotted line on the main figure corresponds to the site problem with $n=2$ with 
$r$ rescaled by the non-universal factor $c_{\rm s}/c_{\rm b}$.}  
\label{fig12_dp}  
\end{figure}

The Monte Carlo results in Figures~\ref{fig11_dp} and~\ref{fig12_dp} for the site 
and bond percolation problems with FBC at the largest size, $L=256$, are in good 
agreement with this scaling behaviour. The insets show the linear variation of 
$\ln\pi_{tn}$ when $r$ is sufficiently large. Different slopes are obtained for 
different values of $n$. As shown in the main figures, after multiplication by 
$n^{-(z+1)}$, a good data collapse is obtained. 

The universality of $\pi_{tn}(r_{\rm eff})$ is illustrated in 
Figure~\ref{fig12_dp}. The dotted line on the main figure corresponds to the site 
problem with $n=2$ after rescaling $r$ by the non-universal factor $c_{\rm 
s}/c_{\rm b}$. The remaining shift between the site and bond data is likely to be 
due to finite-size corrections.


\section{Conclusion}
\label{sec:4}

For the directed percolation problem in the FSS limit, the crossing probability in 
the time direction, $\pi_t$, is a scale-invariant universal function of the effective 
aspect ratio $r_{\rm eff}=ct/L^z$, appropriate for a strongly anisotropic system. 
It also depends on the fraction $f=l/L$ of occupied sites in the initial state,  
where $l$ is the length of the sequence of occupied sites. This function vanishes 
with $f$ as $f^{x_m}$, where $x_m=\beta/\nu$ is the scaling dimension of the order 
parameter. When the $L$ sites are occupied with probability $p_i>0$ at $t=0$, the 
crossing probability is asymptotically the same as for a fully occupied system in 
the initial state. 

Anisotropic scaling, together with the generalization of a simple geometrical 
argument due to Aizenman for isotropic percolation, leads to the scaling behaviour 
for the probability $\pi_{tn}$ to find $n$ critical directed percolation clusters 
surviving at time $t$ on a system with size $L^d$. The probability $\pi_{tn}$ is a 
scale-invariant universal function which decays exponentially with $n^{1+z/d}r_{\rm 
eff}$. The numerical data in $1+1$ dimensions support the conjectured expression 
although further work is needed with larger system sizes, in order to be closer to 
the true asymptotic behaviour. The scaling behaviour should be also examined in 
higher dimensions. In both cases one should use a more efficient simulation method, 
like the ``go with the winner'' strategy introduced by 
Grassberger~\cite{grassberger00,grassberger02,shchur02}.
It would be particularly interesting to study the expected crossover of $\pi_{tn}$, 
due to the proliferation of clusters, at and above the upper critical 
dimension~\cite{shchur02}. It should be easier to do this for directed percolation, 
where $d_{\rm c}=4$, than for isotropic percolation, where $D_{\rm c}=6$.

The present work can be also extended by considering the crossing probability on 
same-spin clusters in strongly ani\-so\-tro\-pic spin systems at equilibrium, like 
the three-dimensional uniaxial ANNNI model at its Lifshitz point, for which the 
critical parameters are now accurately known~\cite{pleimling01}.


\begin{thebibliography}{0}
\bibitem{r} Following a suggestion of the referee, the aspect ratio $r$ is defined 
here as $t/L^z$, which is the inverse of the definition previously used 
in~\cite{turban02} and~\cite{turban03}. With this more natural definition $r$ 
becomes a rescaled time of survival. The non-universal constant $c$ is also changed 
into its inverse.
\end{thebibliography}
\end{document}